\documentstyle[prb,aps,twocolumn]{revtex}
\input epsfig.sty

\begin{document}

\advance\textheight by 0.2in
\twocolumn[\hsize\textwidth\columnwidth\hsize\csname@twocolumnfalse%
\endcsname

\draft
\begin{flushright}
{\tt to appear in Phys. Rev. Letters }
\end{flushright}

\title{Charge glass in two-dimensional arrays of capacitively coupled grains with random offset charges.}
\author{Enzo Granato}
\address{Laborat\'{o}rio Associado de Sensores e Materiais, 
Instituto Nacional de Pesquisas Espaciais, \\ 
12201 - S\~{a}o Jos\'{e} dos Campos, SP, Brazil}
\author{J.M. Kosterlitz}
\address{Department of Physics, Brown University, \\
Providence, RI 02912, USA}
\maketitle

\begin{abstract}
We study the effect of random offset charges in the insulator to conductor
transition in systems of capacitively coupled grains, as realized in
two-dimensional arrays of ultrasmall Josephson junctions.  In presence
of disorder, the  conductive transition and charge ordering at nonzero
gate voltages are both destroyed for any degree of disorder $\sigma$ at
finite temperatures $T$, in the thermodynamic limit, but crossover effects
will dominate at length scales smaller than $\xi _{\sigma ,T}=T^{-\nu}
f(\sigma T^{-\nu })$ , where $\nu \ $ is the thermal critical exponent of
the zero-temperature charge glass transition.  The conductance is linear
and thermally activated  but nonlinear behavior sets in at a crossover
voltage which decreases  as temperature decreases.  For large disorder,
the results are supported  by Monte Carlo dynamics simulations of a
Coulomb gas with  offset charges and are consistent with the thermally
activated behavior found in recent experiments.
\end{abstract}


]

Two-dimensional arrays of Josephson junctions where the charging energy
of the grains is much larger than the Josephson coupling energy are
interesting systems where collective charging effects can be studied
experimentally in great detail \cite{mooij,tinkham,delsing,vdzant}.
In these systems, the net charges in the grains have a long range
logarithmic interaction when the junction capacitance $C$ is small
enough. The charges play the same role as vortices in the resistive
behavior of the array, when capacitive effects can be neglected. This
allows for the possibility of a Kosterlitz-Thouless (KT) charge unbinding transition corresponding to an insulating to conducting  transition
at finite temperature \cite {mooij}. Below the critical temperature
$T_{c}$ , neutral charge pairs would be present in the insulating phase
while above $T_{c},$ screening of the electrostatic potential gives
rise to free isolated charges and a conducting phase. Experimentally, a
transition has in fact been observed in the conductance of small
capacitance arrays of normal and superconducting grains \cite
{mooij,tinkham,delsing} at an apparent temperature consistent with the
estimate based on  logarithmically interacting charges. Recently,
however, a closer analysis of the experimental data of different
groups, revealed that the onset of finite conductance at finite temperatures
could be described just as well by thermal activation of free charges 
\cite{tinkham,delsing} with an activation energy $E_a\approx \frac{1}{4}
E_c$ when the  grains are in the normal state and $E_a\approx \frac{1}{4}
E_c + \Delta $ for superconducting grains, where $2 \Delta$ is the
superconducting energy gap at zero temperature.  This interpretation
neglects the logarithmic interaction between charges expected for
an ideal array \cite{mooij,vdzant} and suggests in turn that, in the
experimental systems,  the interaction is screened by some
unknown mechanism and is essentially short ranged. Although disorder
effects, in the form of random offset charges which may be trapped near
the grain-substrate interface \cite{mooij}, could be a possible
explanation for these results, an understanding of how differently
prepared systems, and therefore different degrees of disorder, can still
lead to the same activated behavior of the linear conductance is still
lacking.  In particular, measurable nonlinear effects resulting from
disorder have not been considered so far.

In this work, we consider the effects of random offset charges on the
charge-anticharge unbinding transition in an otherwise perfect two
dimensional array of capacitively coupled normal or superconducting grains
in the classical regime of large charging energy
$E_{c}$ and finite temperature $T$. We show that, in the presence of offset
charges, the unbinding transition and charge ordering at nonzero gate
voltages are both destroyed for any degree of disorder $\sigma$ in the
thermodynamic limit, consistent with the common behavior of
different experimental arrays. The conductance is linear and thermally
activated  but crossover effects will dominate at length scales smaller
than a disorder and temperature dependent length, $\xi _{\sigma
,T}=T^{-\nu }f(\sigma T^{-\nu })$, where  $\nu \ $ is the thermal
critical exponent characterizing the $T=0$ charge glass
transition.  Nonlinear behavior sets in at a characteristic voltage
$V_{c}\sim T/\xi _{\sigma ,T}$.  For larger disorder, the results are
supported  by a Monte Carlo simulation of the nonlinear conductance
leading to $V_c\sim T^{1+\nu}$ with a very rough estimate $\nu \sim 1.7$.

The electrostatic energy of the net mobile charges $Q_{i}=e^{*}n_{i}$,
multiples of an elementary charge $e^{*}$, located at the sites $i$ of an
array is given by \cite{mooij,fazio} 
\begin{equation}
E={\frac{1}{2}}\sum_{i,j}(Q_{i}+q_{i})C_{ij}^{-1}(Q_{i}+q_{i})
\end{equation}
where $e^{*}=e$, the electron charge, for a normal grain and $e^{*}=2e$
for a superconducting grain. $C_{ij}$ is the capacitance matrix and $q_{i}$
represents a net offset charge in a grain induced by 
charged impurities trapped near the grain-substrate interface
\cite{mooij,wingreen}. The equilibrium properties in the classical
limit are described by the partition function 
$Z=\sum_{\{{n_{i}\}}}e^{-E/kT}$, where the sum is over all integers
$n_{i}$ and the offset charges $q_{i}$ act as quenched disorder with a
nonzero uncorrelated component. We also assume that
tunneling of charges between the grains provides the dynamics to
equilibrate the charge distribution and the junction dissipation is
sufficiently small \cite{mooij}. Since an offset charge equal to an integer number of $e^{*}$, can be
compensated by shifting $n_{i}\rightarrow n_{i}-m$, one can consider
offset charges with a probability distribution of width $\sigma <e^{*}$
peaked at zero. In general, the elements of the inverse capacitance matrix
$C^{-1}_{ij}$ fall off exponentially for $r_{ij}>\Lambda $, where 
$\Lambda \sim \sqrt{ C/C_{g}}$ is a screening length, $C$ the capacitance
between neighboring grains and $C_{g}$ is the capacitance to the
ground. We consider the regime $ r/\Lambda <<1$ where\cite{fazio}
$C_{ij}^{-1}=-{\frac{1}{{2\pi C}}}\log r_{ij}-1/4C$.
Under these conditions, the interaction energy above describes a two
dimensional Coulomb gas of logarithmically interacting charges $Q_{i}$ in a
background of quenched random charges $q_{i}$. This problem has been considered
before in relation to the effects of geometrical disorder in classical
superconducting arrays in a magnetic field where $Q_{i}$ represents vortices
and $q_{i}$ a random flux \cite{gk} and in the gauge-glass model \cite{fisher}
of disordered superconductors where the $q_{i}$ are random but correlated.

In the absence of offset charges, $q_{i}=0$, a charge-anticharge unbinding
transition in the KT universality class has been predicted \cite{mooij} at a finite temperature $kT_{c} \approx {e^{*}}^{2}/8\pi C$.
However, in the presence of offset charges with a nonzero
component of uncorrelated disorder, $<q_{i}q_{j}>\ =\ \sigma ^{2}\delta _{ij}$,
the behavior is drastically changed as bound charge-anticharge pairs are
unstable at a sufficiently large length scale $\xi $ determined by the
degree of disorder. In fact, the net charge fluctuation due to random
charges in a domain of area $\pi L^{2}$ will neutralize a test charge 
$Q=e^{*}$ when $L=\xi _{\sigma }={\ e^{*}}/\sqrt{\pi }\sigma $. For length
scales larger than $\xi _{\sigma }$, charges are unbound and will contribute
to the linear conductance $G$ provided
that charges are mobile. This shows that even for small $\sigma $, disorder
is a strongly relevant perturbation.

In general, there will be also a contribution from correlated
disorder.  However, this is either irrelevant
at the fixed point of the pure system or at most marginal for
sufficiently small disorder.  To lowest order in the wavevector $k$, the
moments of disorder distribution are given by $<q_{k}>=0 $ and
$
<q_{k}q_{-k}>=\sigma ^{2}+\sigma _{2}^{2}k^{2}+O(k^{4})
$.
The first term  corresponds to uncorrelated randomness as
discussed above and the second term  to offset charges correlated
as random dipoles $p=q\ |r_{i}$ $-r_{j}|$ , where $r_{i\text{ }}$and $r_{j}$
are neighboring sites, since $<p_{k}p_{-k}>=\sigma _{2}^{2}k^{2}$ . Higher
order terms correspond to multipole correlations. Using standard methods
to convert the partition function of Eq.(1) to a sine-Gordon model \cite
{kogut} and the replica trick to average the free energy over disorder,
we obtain an effective reduced Hamiltonian in terms of $n$ replicas 
\begin{eqnarray}
H/kT &=&\frac{1}{2}\int d^{2}r[\frac{1}{K}\sum_{\alpha }(\nabla \varphi
^{\alpha })^{2}+(\frac{2\pi \sigma}{e^{*}})^{2}\sum_{\alpha ,\beta }\varphi
_{r}^{\alpha }\varphi _{r}^{\beta }  \nonumber \\
&&+(\frac{2\pi \sigma _{2}}{e^{*}})^{2}\sum_{\alpha ,\beta }\nabla \varphi ^{\alpha
}\nabla \varphi ^{\beta \ }-y\sum_{\alpha }\cos (2\pi \varphi _{r}^{\alpha
}) ],
\end{eqnarray}
where $K=E_c/2\pi^2 kT$. Higher order gradient terms have been neglected.  In the absence
of disorder this model has a massless phase for $\pi K>2$ when the charge
fugacity $y$ is irrelevant corresponding  to an insulating phase. 
For $\pi K<2$, $y$  
is relevant leading to a conducting phase. From renormalization-group
arguments, the relevance of the disorder perturbation on the massless
phase can now be determined from the eigenvalue $\lambda =d-x$ where $
2x $ is correlation function exponent evaluated at the unperturbed
fixed point. For uncorrelated disorder, $\sigma \neq 0$, it is
sufficient to consider the diagonal contribution $\alpha =\beta $. This
has an eigenvalue $ \lambda =2$ in the massless phase since
$x=(d-2)=0$ in two dimensions, which corresponds to a scaling behavior
$\sigma ^{\prime }\ ^{2}=\sigma ^{2}\ b^{2} $ under a change of lattice
spacing by a factor of $b$, in agreement with the domain argument
given above leading to a characteristic disorder length $\xi _{\sigma
}\propto 1/\sigma $. For correlated disorder
alone, $\sigma _{2}\neq 0$, the contribution from the
diagonal term $\sum_{\alpha }\nabla \varphi _{\alpha }\nabla \varphi
_{\alpha \ }$ just renormalizes the value of $K$ but the offdiagonal
perturbation $\sum_{\alpha \neq \beta }\nabla \varphi _{\alpha }\nabla
\varphi _{\beta \ }$ is marginal, $\lambda =0$ . The
higher order gradient terms in the Hamiltonian are then irrelevant.

The disorder dependent length $\xi _{\sigma }$ is the scale up to which the
charges $Q_{i}$ are mobile. Beyond this scale, they are neutralized by the
random background of quenched charges $q_{i}$. At a scale $L>\xi
_{\sigma }$, the system is better described as a set of pinned charges
interacting by a short range potential and we expect that it behaves
similarly to a charge glass. Since these charges play the same role as
vortices in a two-dimensional superconductor, one expects that glass order
occurs at $T=0$ only, with a thermal correlation length $\xi_T
\propto T^{-\nu }$ where $\nu $ is a thermal critical exponent
associated with the $T=0$ fixed point \cite{fisher}. Assuming that the model
of Eq.(1) is in the same universality class, we expect that $\nu$ 
$\gtrsim 2$. Standard crossover scaling leads to a correlation length $
\xi _{\sigma ,T}=T^{-\nu }f(\sigma T^{-\nu })$ where the scaling
function $f(x)$ is a function of the dimensionless ratio $x=\xi _{T}/\xi
_{\sigma }$ with $f(x)$ $\sim 1$ for $x>>x_{c}$ and $f(x)$ $\sim
1/x$ for $x<<x_{c}$ so that $\xi _{\sigma ,T}\propto 1/\sigma $ when $
\sigma T^{-\nu }<x_{c}$ and $\xi _{\sigma ,T}\propto T^{-\nu }$
for $\sigma T^{-\nu }>x_{c}$ . For an infinite system, the predicted
insulating -conducting transition in the absence of disorder is destroyed by
any finite density of uncorrelated offset charges but crossover effects may
lead to an apparent transition if $\xi _{\sigma ,T}$ is larger than the
system size. Note that the above arguments are not strictly correct in the
limit $\sigma =0$ or $T=0$ as it does not allow for the possibility of a
transition to a true insulating phase of bound pairs of charges at finite
$T$ when $\sigma =0$ or for quantum tunneling \cite{delsing} at $T=0$.
In principle, such effects could be incorporated in the crossover scaling
analysis at the expense of some major extra complications but experiments 
\cite{mooij,tinkham,delsing} indicate that there is a range of temperatures
over which purely classical behavior is observed and that behavior
characteristic of the absence of disorder is never observed. The crossover
scaling analysis outlined above should be relevant to experimental systems
at temperatures above the crossover temperature where quantum tunneling
dominates \cite{delsing}.

In addition to $\xi $, another length scale, $\xi_{V}$,
is set by an applied finite voltage $V$ across the system\cite{current}.
The additional contribution to the energy has the form $\sum_{i}Q_{i}Ex_{i}$
for an uniform electric field $E$ in the $x$ direction. Thermal fluctuations
alone, of typical energy $kT$, leads to a characteristic length $\xi
_{V}\sim kT/e^{*}E$ over which single charge motion is possible. For $\xi
_{V}>\xi $, when charges are unbound, this leads to a linear response to
the applied voltage and a finite conductance limited by the tunneling rate
across each junction in the array. For $\xi _{V}<\xi $, the contribution
from the applied voltage must be balanced against the interaction energy
with other charges leading to nonlinear behavior. A crossover from
linear to nonlinear behavior occurs when these two lengths are comparable,
i.e., at a characteristic voltage $V_{c\ }{\sim T/}\xi _{\sigma ,T}{\ =\ }
T^{\nu \ +\ 1}\ g(\sigma T^{-\nu })$, where $g(x)=1/f(x)$ .

The finite correlation length $\xi $ in presence of disorder also determines
the behavior of the activation energy in the regime of linear conductance. For
vanishing voltage, the current arises from thermal dissociation of the most
weakly bound charge pairs. For normal metals, the energy necessary to
unbind a typical pair is $E=\frac{E_{c}}{\pi }\log (\xi /a)+E_{o}$,
where $E_{o}\simeq E_{c}/2$ is the core energy of a pair with one lattice
spacing separation. For sufficiently large disorder such that $\xi \sim a$,
the conductance should then have an Arrhenius behavior with an activation
energy given \cite{tinkham} by $E_{a}=E_{o}/2$. However, for moderate
disorder the activation energy has a temperature and disorder dependent
logarithmic correction $\log (\xi _{\sigma ,T}/a)$. In the pure case where $
\sigma =0$ and $\xi _{\sigma ,T}$ $\rightarrow $ $\infty $, this implies an
infinite activation energy and the linear conductance vanishes consistent
with $V_{c\ }{\sim T/}\xi _{\sigma ,T}\rightarrow 0$. The logarithmic
dependence on $\xi _{\sigma ,T}$ of the activation energy also indicates
that its value is only drastically changed for small disorder or low
temperatures where $\xi _{\sigma ,T}$ is large but is roughly insensitive to
larger values of disorder or temperature. For superconducting grains, the
arguments above still apply except that an additional energy \cite{tinkham}
equal to the superconducting energy gap $E_{g}$ is required to create
quasiparticles which allow charges on the grains to tunnel to
neighboring grains, in the absence of Josephson tunneling.

In the presence of a gate voltage $V_{g}$ on each grain, in the absence of 
disorder there will be an
induced charge $C_{g}V_{g}$ on each grain acting as a frustration parameter 
\cite{mooij}$f=C_{g}V_{g}/e^{*}$ analogous to the flux quanta per plaquette
in Josephson junction arrays in a magnetic field. The properties of the
system should be periodic in $f$ with period $1$. In particular, the Coulomb
gap width oscillates as a function of $V_{g}$ with period $e^{*}/C_{g}$,
often used as an estimate of $C_{g}$. For fractional values of $f$ the
ground state is a charged ordered state with an average number $f$ of elementary
charges per grain. For rational $f=p/q$, this leads to a ground
state with a unit cell of size $q\times q$ and a discrete degenerate ground
state \cite{teitel}. A finite temperature transition is expected from the
charge ordered phase to a disordered phase in addition to the
charge-anticharge unbinding transition. In presence of disorder, the
behavior at integer $f$ will be the same as $f=0$, i.e., at any finite
temperature the interacting charge system is disordered and the conductance
is finite. This behavior should persist for fractional values of $f$
since, in addition to the discrete symmetry excitations such as domain walls,
thermally excited net charges interacting logarithmically are also present
which will be screened by disorder at length scales larger than $\xi $
leading to a linear thermally activated conductance. However, one still
may question if charge order remains stable in presence of disorder. It is
sufficient to consider the case $f=1/2$ since the ground states for $f<1/2 $
are less stable \cite{teitel}. At $f=1/2$ and in the absence of disorder, it
is known that the ground state \cite{teitel} consists of $2\times 2$ unit
cells with an antiferromagnetic arrangement of $n_{i}-f=\pm 1/2$ net
fractional charges, in units of the elementary charge $e^{*}$ . The lowest
energy excitations  are domain walls separating the two degenerate
ground states, $\pm 1/4$ charges at domain wall corners and unit excess
charges interacting logarithmically. In a coarse-grained description, the
variables in a unit cell can be replaced by Ising variables $S_{r}=\pm 1$,
which describe the charge ordered state, and charges $Q_{R}=\sum S_{r}/4$ averaged over the unit cell \cite{knops}.
This leads to an effective Hamiltonian \cite{gk}
\begin{eqnarray}
H/kT &=&L_{1}\sum_{<r,r\prime >}S_{r}S_{r^\prime }+2\pi ^{2}K\sum_{R,R^\prime
}Q_{R}\ G(R-R^\prime )\ Q_{R^\prime }  \nonumber \\
&&+L_{2}\sum_{r}S_{r}q_{r}+4\pi ^{2}K\sum_{rR}Q_{R}G(R-r)q_{r}
\end{eqnarray}
where $G(r)=\log (r/a)+\pi /2$. The Ising variables $S_{r}$ are coupled
antiferromagnetically by a nearest neighbor interaction $L_{1}$ and to a random field proportional to $q_{r}$. The excess
charges $Q_{R}$ are still coupled to the random offset charges $q_{r}$ and
will therefore lead to finite conductance as discussed before. Since the
offset charges coupled as random fields to the Ising variables, charge order
will be destroyed even for small disorder. There will be
however a characteristic length scale \cite{moore} $\xi _{\sigma }^{\prime
}\sim \exp (c/\sigma ^{2})$, beyond which random offset charges destroy the
charge ordered state. For sufficiently small $\sigma $, $\xi_{\sigma}^{'} >
\xi _{\sigma }$ and there can be an
intermediate length scale where charge order remains but excess charge is
effectively unbound. Similar results should apply for $f<1/2$ and we expect
that disorder destroys charge ordering for any finite gate voltage.

We have studied the linear and nonlinear conductance of the Coulomb gas
in Eq.(1) numerically using classical Monte Carlo nonequilibrium
dynamics\cite{fisher}, regarding the Monte Carlo time as real time.
This should be a reasonable approximation in the overdamped limit and
when quantum effects can be ignored. It also  assumes that the
tunneling rate which provides the equilibrium charge distribution
satisfies the global rule\cite{geig} when it depends on the energy
difference of the whole array. In the simulations, a Monte Carlo step
consists of adding a dipole of unit charges and unit length to a nearest
neighbor charge pair, $(n_{i,}n_{j})$, corresponding to a tunneling  of
the positive or negative charge by a unit of length. The external
electric field biases the added dipole, leading to a
current as the net flow of charge is in the direction of the electric field.
The unit of time corresponds to a complete Monte Carlo pass
through the system.  The linear conductance $G_L=\lim_{E\rightarrow 0}
J/E$ is obtained from the equilibrium  current fluctuations, without
imposing a voltage bias, from the fluctuation-dissipation relation $G_L
=\frac{1}{2 kT} \int dt <I(0) I(t) > $.

\begin{figure}[tbp]
 \centering\epsfig{file=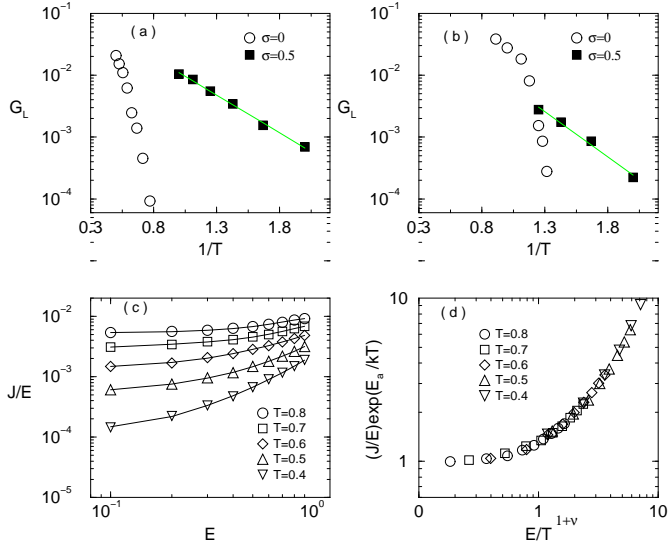,bbllx=6cm,bblly=0.7cm,bburx=16cm,
 bbury=18cm,width=4.5cm}
\caption{(a) Linear conductance $G_L$ with zero gate voltage ($f=0$) and
(b) with nonzero gate voltage ($f=1/2$). The straight lines indicate an Arrhenius behavior. Temperature $T$ is measured units of $E_c/2\pi^2$. 
(c) Nonlinear conductance $J/E$ as a function of temperature, 
for $\sigma = 0.5$ and ($f=0$). (d) Scaling plot of the nonlinear
conductance in (c). }
\end{figure}

Figs.(1a,b) show the behavior of the linear conductance $G_L$ in the
absence of disorder, $\sigma =0$, and in the presence of a (large)
random Gaussian distribution of offset charges with a standard
deviation $\sigma =0.5$. Since large disorder leads to small $\xi$, we
can use a small system size  $L \times L$ with $L=16$. When $\sigma =0$,
the conductance appears to vanish at $T_{c}\approx 1.4$ for
$f=0$ and $\approx 0.8$ for $f=1/2$, corresponding  to the critical
temperature for the conductive  transition for zero and nonzero gate
voltages. However, for $\sigma =0.5$ , the conductance remains finite
much below the critical temperature for the pure system.  In fact, as
indicated in the figure, the data in the low temperature range can be
reasonably fitted to the Arrhenius activated form
$G=G_{o}e^{-E_{a}/kT}$ in agreement with a disordered phase. Moreover,
the calculation of $G_L$ from current fluctuations ensures that the
activated behavior is not an artifact of a voltage bias since it is
obtained without an external electric field.  In experiments, however,
a voltage bias is often used and nonlinear effects should be taken into
account.  In Fig.(1c) we show the behavior of the nonlinear conductance
as a function of the applied electric field and temperature for $\sigma
=0.5$ and $f=0$. For small $E$, $J/E$ tends to the constant finite
value of Fig.(1a) which decreases with temperature. For the highest
temperature, $ T=0.8$, the range of $E$ in which this behavior is
apparent is more pronounced. For increasing $E$ it clearly crosses over
to a nonlinear behavior in agreement with the analysis leading to a
crossover voltage $V_{c}$ which is both temperature and disorder
dependent. The nonlinear conductance can be cast into a scaling form
\cite{fisher}, $J/E G_L = F(E/E_c)$, where $F$ is some scaling function
with $F(0)=1$. For large disorder we have  $E_c \sim T^{\nu + 1}$, and
a scaling plot allows an estimate of $\nu$ .  This  is shown in Fig.(1d)
where a reasonable  data collapse is  obtained by adjusting $E_a$
and $\nu$ for the largest temperatures, giving $\nu \sim 1.7$, in rough
agreement with the $T=0$ stiffness exponent for gauge glass models\cite{fisher}. This implies that the characteristic voltage $V_{c}$ where nonlinearities set in
should exhibit a nontrivial power law $V_c \sim T^{\nu +1}$ and should
be accessible experimentally.

\medskip 

This work was supported by FAPESP,  Proc. 97/07250-8 (E.G.) and  Proc. 
97/9188-8 (J.M.K.), and by a joint CNPq-NSF grant. We acknowledge the support
from ICTP, where part of the work was done.


\begin{references}
\bibitem{current} This is the analogous effect of an applied current 
for vortex motion  in superconductors \cite{fisher}.
\end{references}
\end{document}